\documentclass[prd,twocolumn,amsmath,amssymb,floatfix, superscriptaddress]{revtex4}

\usepackage{bm}
\usepackage{amsmath}
\usepackage{epsf}
\usepackage{color}
\usepackage{natbib}
\usepackage{graphicx}
\def\simlt{\lesssim}
\def\simgt{\gtrsim}

\def\be{\begin{equation}}
\def\ee{\end{equation}}
\def\ba{\begin{eqnarray}}
\def\ea{\end{eqnarray}}
\def\nn{\nonumber}

\def\bigoh{{\mathcal O}}
\def\ellmax{\ell_{\rm max}}
\def\n{\widehat{\bf n}}

\newcommand{\vl}{{\bm \ell}}

\newcommand{\wj}[6]{\left(
                           \begin{array}{ccc}
        \! #1\! & #2\!  & #3\!  \\
        \! #4\! & #5\!  & #6\!
                           \end{array}
                   \right)}

\begin{document}

\title{${\bm B}$-mode CMB Polarization from Patchy Screening during Reionization}

\author{Cora Dvorkin}
\affiliation{Department of Physics, University of Chicago, Chicago IL 60637}
\affiliation{Kavli Institute for Cosmological Physics and Enrico Fermi Institute, University of Chicago, Chicago IL 60637, U.S.A.}

\author{Wayne Hu}
\affiliation{Department of Astronomy \& Astrophysics, University of Chicago, Chicago IL 60637}
\affiliation{Kavli Institute for Cosmological Physics and Enrico Fermi Institute, University of Chicago, Chicago IL 60637, U.S.A.}

\author{Kendrick M. Smith}
\affiliation{Institute of Astronomy, University of Cambridge, Cambridge, CB3 0HA, UK}

\begin{abstract}
\baselineskip 11pt
$B$-modes in CMB polarization from patchy reionization arise from two effects:
generation of polarization from scattering of quadrupole moments by reionization bubbles,
and fluctuations in the screening of $E$-modes from recombination.
The scattering contribution has been studied previously, but the screening contribution has not
yet been calculated.
We show that on scales smaller than the acoustic scale ($\ell\simgt 300$),
the B-mode power from screening is larger than the B-mode power from 
scattering.
The ratio approaches a constant $\sim$2.5 below the damping scale ($\ell\simgt 2000$).
 On degree scales relevant for gravitational waves ($\ell\simlt 100$), screening $B$-modes
have a white noise tail and are subdominant to the scattering effect.
These results are robust to uncertainties in the modeling of patchy reionization.
\end{abstract}
\maketitle
\section{Introduction}
 In linear theory, the presence of a curl or $B$-mode pattern in the polarization of the
 CMB is an indication that there are contributions from gravitational waves or vector
 perturbations that, unlike density perturbations, can impart a sense of handedness to the polarization.   Beyond linear theory,
 it is well known that second order effects such as gravitational lensing can produce
 $B$-modes from density fluctuations \cite{ZalSel98}.   In fact, any process that modulates the polarization 
 amplitude, direction, or position on the sky can generate $B$-modes from the intrinsic
 $E$-modes at recombination
(e.g.~\cite{HuHedZal02}).

In this {\it Brief Report}, we study one such contribution: the amplitude modulation due to
patchy screening of the primary polarization due to inhomogeneous reionization.  
Whereas the analogous effect of patchy generation of polarization from scattering
of the quadrupole anisotropy during reionization has been well-studied in the literature
\cite{Hu00,weller99,Liuetal01,Sanetal03,MorHu06,Doretal07},
patchy screening is usually neglected as a small contribution.    While negligible at large angles,
we show here that the two effects are always comparable in magnitude beyond the damping
tail with $\sim 2.5$ times as much power in the screening effect. 

\section{Patchy Screening}
\subsection{Formalism}

Scattering of CMB radiation during reionization out of the line of sight suppresses the primary
temperature and polarization anisotropy from recombination as $e^{-\tau}$ where $\tau$
is the Thomson optical depth.   If $\tau$ varies across the line of sight this suppression itself introduces anisotropy as
\ba
T(\n)&=&e^{-\tau(\n)}T^{({\rm rec})}(\n) \,, \nn \\
(Q\pm iU)(\n) &=&e^{-\tau(\n)}  (Q \pm iU)^{({\rm rec})}(\n)  \,,
\ea
where $T$ is the temperature fluctuation and $Q$ and $U$ are the polarization Stokes parameters,
and we have assumed that the dominant temperature fluctuations from recombination are below the
angular scale subtended by the horizon during reionization.
These recombination fluctuations can be decomposed in multipole moments in the
usual way with the additional assumption that the polarization contains
$E$-modes only
\ba
T^{({\rm rec})}(\n) &=& \sum_{\ell m} T_{\ell m}^{({\rm rec})} Y_{\ell m}(\n) \,, \nn\\
(Q\pm iU)^{(\rm rec)}(\n) &=& -\sum_{\ell m} (E_{\ell m}^{({\rm rec})} \pm i  B_{\ell m}^{({\rm rec})} )
 [ {}_{\pm 2}Y_{\ell m}(\n) ] \nn\\
 &=&  -\sum_{\ell m} E_{\ell m}^{({\rm rec})}  [ {}_{\pm 2}Y_{\ell m}(\n) ] \,.
\ea
Decomposing the anisotropic part of the optical depth into multipole moments
\ba
\tau(\n) &=& \bar\tau + \sum_{\ell\ge 1} \sum_m \tau_{\ell m} Y_{\ell m}(\n) \,,
\ea
we  obtain in the  $(\tau(\n) -\bar\tau) \ll 1$ limit
\ba
T_{\ell m}^{({\rm scr})} & = &  - e^{-\bar \tau}
\sum_{\ell' m' \ell'' m''}\tau_{\ell''m''}T_{\ell'm'}^{({\rm rec})}  \\
&& \times (-1)^m\sqrt{2\ell+1} W_{\ell\ell'\ell''}^{000} \wj{\ell}{\ell'}{\ell''}{-m}{m'}{m''}  \,, \nn\\
E_{\ell m}^{(\rm scr)} &=& - e^{-\bar \tau}
   \sum_{\ell'm'\ell''m''} \tau_{\ell''m''}  E_{\ell'm'}^{(\rm rec)}e_{\ell \ell'\ell''}  \nn\\
  &&\times (-1)^m \sqrt{ 2\ell+1 } W_{\ell\ell'\ell''}^{220}  \wj{\ell}{\ell'}{\ell''}{-m}{m'}{m''} \,, \nn  \\
i B_{\ell m}^{(\rm scr)} &=&
  - e^{-\bar \tau}\sum_{\ell'm'\ell''m''} \tau_{\ell''m''}E_{\ell'm'}^{(\rm rec)} o_{\ell \ell'\ell''}  \nn\\
  &&\times (-1)^m \sqrt{ 2\ell+1 } W_{\ell\ell'\ell''}^{220}    \wj{\ell}{\ell'}{\ell''}{-m}{m'}{m''} \,, \nn
\ea
where $e_{\ell \ell' \ell''}=[1+(-1)^{\ell+\ell'+\ell''}]/2$ and
$o_{\ell \ell' \ell''} = [1-(-1)^{\ell+\ell'+\ell''}]/2$ picks out even and odd triplets and
\ba
W_{\ell \ell'\ell''}^{s s' s''} = \sqrt{{(2\ell'+1)(2\ell''+1)\over4\pi}} \wj{\ell}{\ell'}{\ell"}{-s}{s'}{s''}\,.
\ea
Here we have given the contribution to the anisotropy from 
the $\ell > 0$ terms in $e^{-\tau}$:
\be
X_{\ell m}^{\rm (scr)} = X_{\ell m} - \langle e^{-\tau} \rangle X_{\ell m}^{\rm (rec)}  \,,\label{eq:scr_definition}
\ee
where $X \in \{ T,E,B \}$.
The additional power spectra contributions
\be
\langle X_{\ell m}^* Y_{\ell'm'} \rangle = \delta_{\ell \ell'} \delta_{m m'} C_\ell^{XY} \,,
\ee 
become
  \cite{HuHedZal02}
\ba
C_\ell^{TT ({\rm scr})}&=& e^{-2\bar\tau}\sum_{\ell'\ell''}C_{\ell''}^{\tau\tau}C_{\ell'}^{TT ({\rm rec})}
(W_{\ell \ell'\ell''}^{000} )^2\,,  \nn\\
C_\ell^{TE ({\rm scr})}&=&e^{-2\bar\tau} \sum_{\ell'\ell''}C_{\ell''}^{\tau\tau}C_{\ell'}^{TE ({\rm rec})}
(W_{\ell \ell'\ell''}^{000}W_{\ell \ell'\ell''}^{220} )\,,  \nn\\
C_\ell^{EE ({\rm scr})} & =& e^{-2\bar\tau} \sum_{\ell'\ell''}C_{\ell''}^{\tau\tau}C_{\ell'}^{EE ({\rm rec})}
e_{\ell\ell'\ell''}
(W_{\ell \ell'\ell''}^{220} )^2\,, \nn\\
C_\ell^{BB ({\rm scr})} & =& e^{-2\bar\tau} \sum_{\ell'\ell''}C_{\ell''}^{\tau\tau}C_{\ell'}^{EE ({\rm rec})}
o_{\ell\ell'\ell''}
(W_{\ell \ell'\ell''}^{220} )^2\,.  \label{eq:cl_src}
\ea
We evaluate the recombination power spectra by setting  $\tau=0$.  These sums can be 
efficiently evaluated in position space as shown in the Appendix.

Note that in the decomposition defined in Eqn.~(\ref{eq:scr_definition}),
the isotropic screening term gains a contribution from the rms fluctuations in $\tau$
\begin{equation}
 \langle e^{-\tau} \rangle^2 \langle X_{\ell m}^{\rm (rec)*} Y_{\ell' m'}^{\rm (rec)}\rangle
 \approx \delta_{\ell\ell'}\delta_{mm'} e^{-2\bar\tau}(1+\tau_{\rm rms}^2) C_\ell^{XY ({\rm rec})} \nn
 \end{equation}
 where
 \begin{equation}
 \tau_{\rm rms}^2 = \sum_{\ell \ge 1} {2 \ell +1 \over 4\pi} C_\ell^{\tau\tau}\,.
 \end{equation}
 This term is ordinarily not included in the standard calculation but is a correction that is
 second order in $(\tau(\n)-\bar\tau)$.

\subsection{${\bm B}$-Mode Scaling}

We now focus on the $B$-mode generation from screening
and its relationship to other secondary effects.
To extract the large and small angle scaling behavior of the $B$-modes it is useful 
to take the flat-sky approximation where Fourier moments replace harmonic coefficients
\be
(Q\pm iU)(\n) = \int {d^2 {\vl} \over (2\pi)^2}  [E({\vl}) \pm i B({\vl})] e^{\pm 2 i\phi_\vl} e^{i
{\vl} \cdot {\n}}\,,
\ee
where $\phi_{\vl}$ denotes the angle between ${\vl}$ and the axis on which $Q$ is defined.
The power spectrum
\be
\langle B^*({\vl}) B({\vl}') \rangle = (2\pi)^2 \delta({\vl}-{\vl}') C_\ell^{BB} \,,
\ee
then becomes for the screening modes
\be
C^{BB ({\rm scr})}_\ell =e^{-2\bar\tau} \int {d^2{{\vl}'}\over (2\pi)^2}C^{EE ({\rm rec})}_{{\ell}'} C^{\tau\tau}_{|{\vl}-{\vl}'|}\sin^2{(2\phi_{{\vl'}})} \,,
\ee
where we have set $\phi_{\vl}=0$ for convenience.

The $B$-mode power spectrum therefore represents a convolution of recombination and
$\tau$ power spectra.  Its main properties
are defined by the well-determined shape of the primary
$E$-mode power spectrum.   In particular, the primary $E$-modes have little power above
the acoustic scale
$\ell <  \ell_A \approx 300$ and below the damping scale $\ell >  \ell_D \approx 2000$.
In these two limits, the screening $B$-modes take on a particularly simple form.
For $\ell \ll \ell_A$,
\be
C^{BB ({\rm scr})}_\ell \approx {e^{-2\bar\tau}  \over 2} \int {d^2{{\vl}'}\over (2\pi)^2}C^{EE ({\rm rec})}_{{\ell}'} C^{\tau\tau}_{{\ell}'}={\rm const.}
\ee
is white noise in form.   Note that this property is independent of the amount of
low $\ell$ power in $C_\ell^{\tau\tau}$.  Only optical depth fluctuations on scales comparable to
$\ell_A$ can modulate the polarization into large angle $B$-modes.  

For $\ell \gg \ell_D$, the $B$-modes reflect the shape of the
$\tau$ power spectrum
\be
C^{BB ({\rm scr})}_\ell \approx {1  \over 2}   E_{\rm rms}^2{C^{\tau\tau}_\ell} e^{-2\bar\tau}
\label{eqn:smallscale}
\ee
with an amplitude that is determined by  the total power in the primary $E$-modes
\be
E_{\rm rms}^2 = \int {d^2 {\vl} \over (2\pi)^2} C^{EE ({\rm rec})}_{{\ell}}  \approx \sum_\ell {2\ell+1 \over 4\pi}
C^{EE ({\rm rec})}_{{\ell}} \,.
\ee
Under the damping tail,
 the primary $E$-modes are much smoother than $\tau$ fluctuations and so $E_{\rm rms}$ 
 simply represents the typical level of $E$-mode that is modulated into small scale $B$-mode polarization
 by $\tau$.

It is useful to compare these scalings with those of the two other well-known sources of secondary
$B$-modes: gravitational lensing and patchy scattering of quadrupole anisotropy at reionization.
 Lensing obeys a similar form with deflection angles playing the role of $\tau$
 and polarization gradients playing the role of $E$.  In particular, it has a white noise
 form for $\ell \ll \ell_A$ and follows the deflection power spectrum for $\ell \gg \ell_D$ with an amplitude
 set by the total gradient power in the primary polarization \cite{ZalSel98,Hu00b}.  
 
 For patchy scattering of the quadrupole fluctuations during reionization 
 \cite{Hu00}
 \be
 C^{BB ({\rm sca})}_\ell \approx {3 \over 100}  {C^{\tau\tau}_\ell} Q_{\rm rms}^2 e^{-2\tau_{\rm eff}}
 \ee
 for scales much smaller than the reionization bump $\ell \gg \ell_R \approx 20$.  
 Here $Q_{\rm rms}$ is the rms quadrupole at reionization and $\tau_{\rm eff}$ is the typical optical depth between $z=0$ and the scatterers,
 so that $\tau_{\rm eff} \simlt \bar\tau$.   The crucial difference
 between this effect and the patchy screening effect is that the quadrupole is coherent on
 large angles $\ell_R \ll \ell_A$ and so the $B$-modes are in the small scale limit for all relevant
 scales. 
 
 In particular for $\ell\gg \ell_D$,
 \be
{C_\ell^{BB ({\rm scr})}\over C_\ell^{BB ({\rm sca})}} \approx {50\over3}{E_{\rm rms}^2\over Q_{\rm rms}^2}e^{2(\tau_{\rm eff}-\bar\tau)}\,.
\ee
For the WMAP5 cosmology \cite{Komatsu:2008hk}, $\Omega_b h^2=0.02265$, 
$\Omega_m h^2=0.137$, $h=0.701$, $\Omega_\Lambda=0.721$, 
$\bar \tau=0.084$, $A_{S}=2.16 \times 10^{-9}$, $n_s=0.96$, the rms fluctuations are
 $E_{\rm rms}=6.9 \mu$K and $Q_{\rm rms}=17.9 \mu$K and so the ratio 
of power approaches a factor of $\sim 2.5$.  Independently of the form of $C_\ell^{\tau\tau}$, the screening
$B$-modes dominate the scattering ones on scales near the damping tail and below.  

Conversely, independently of the power in $C_\ell^{\tau\tau}$ at multipoles $\ell<\ell_A$, the 
screening $B$-modes fall as white noise whereas the scattering ones can continue to
rise if there is large-scale power in $\tau$.  
Therefore only the scattering $B$-modes are relevant as contamination to the gravitational
wave signal at $\ell \sim 100$.   

\begin{figure}[htbp] 
\begin{center}
\includegraphics[width=3.5in]{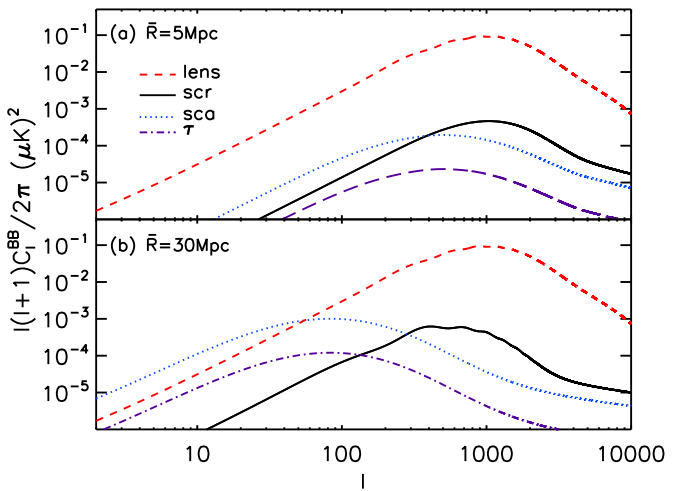}
\end{center}
\caption{$B$-mode power spectra. Three different contributions are being plotted: lensed $B$-modes, patchy $B$-modes from Thomson scattering, and patchy $B$-modes from screening.  The model is the WMAP5 cosmology with (a) the fiducial reionization model of 
$\{\bar \tau, \Delta_y, b, \bar{R}, \sigma_{\ln R}\}=\{ 0.084, 19.0, 6.0, 5\mbox{ Mpc}, \ln(2) \}$, and 
(b) a maximal model with the same parameters except $\bar{R}=30$ Mpc.  For reference the $\tau$ power spectrum in dimensionless units is also plotted.}
\label{fig:Bmodes_fiducial}
\end{figure}

We show examples of these behaviors in Fig.~\ref{fig:Bmodes_fiducial}.
We take two representative models, a ``fiducial'' model in which $C_\ell^{\tau\tau}$ has only small
scale power contributed by relatively small ionization patches, and a ``maximal'' model
which has substantial power at $\ell \sim 100$ from patches that are substantially
larger than those expected in current models of reionization.  
The large-bubble model maximizes the scattering $B$-mode contamination to the gravitational
wave signal \cite{MorHu06}.

More precisely, in both models the ionization history $x_e(z)$ is the same as the fiducial model in \cite{Dvorkin:2008tf}, where the parameter $\Delta_y$ controls the duration of partial ionization. Specifically, our choice of $\Delta_y=19$ gives a range of $6\lesssim z\lesssim 14$. The total optical depth to recombination is taken to be $\bar \tau=0.084$.
During partial reionization, the ionized regions are represented by spherical bubbles with a log-normal distribution 
\cite{Furlanetto:2004nh,Zahn:2006sg} with characteristic size given by $\bar{R}$ (in Mpc) and distribution width given by $\sigma_{\ln R}$.
We take $\{\bar R, \sigma_{\ln R}\} = \{5,\ln(2)\}$ in the fiducial model, and $\{\bar R, \sigma_{\ln R}\} = \{30,\ln(2)\}$ in the large-bubble model.
We assume that the bubbles are linearly biased tracers of the large-scale matter density field, with bias $b=6$ using the construction of Ref.~\cite{WanHu05}.
Note that rms fluctuations in $\tau$ for the small and large-bubble models
are $\tau_{\rm rms}=0.007$ and $\tau_{\rm rms}=0.017$ and so both satisfy the condition that
$\tau_{\rm rms} \ll \bar \tau$.

For reference, the small scale limits to the other power spectra for patchy screening can
be obtained by replacing $E_{\rm rms}^2/2$ with $T_{\rm rms}^2$ for
temperature  in Eqn.~(\ref{eqn:smallscale}) where
\ba
T_{\rm rms}^2  &\approx& \sum_\ell {2\ell+1 \over 4\pi}
C^{TT ({\rm rec})}_{{\ell}} \,,
\ea
whereas the $E$ polarization and $B$ polarization power are equal.
The $TE$ cross power spectrum and correlation coefficient is zero in this limit.

\section{Discussion}

Whereas the patchiness of reionization itself remains theoretically uncertain and observationally
unconstrained, its effects on CMB polarization are well defined.  In addition to the well-known
patchy scattering of the quadrupole moment, the patchy screening of primary polarization 
produces a $B$-mode.  

 Independently of the form of the patchiness, this contribution is larger
than the scattering effect by a factor of $\sim 2.5$ in power on scales below the damping tail.   Conversely, on scales above the acoustic scale, the patchy screening effect always falls
as white noise unlike the scattering effect.  Consequently screening $B$-modes are relatively
more important when considering reionization contributions at arcminute scales but are
largely unimportant for the degree scales relevant for gravitational wave studies.

\acknowledgments 
We would like to thank Anthony Challinor for useful discussions.
CD and WH were supported by the KICP through the grant NSF PHY-0114422 and
the David and
Lucile Packard Foundation. 
WH was additionally supported by  the DOE through 
contract DE-FG02-90ER-40560. 
KMS was supported by an STFC Postdoctoral Fellowship.
\clearpage

\appendix

\section{Efficient Position-space Forms}

The harmonic-space forms for the power spectra given in Eqn.~(\ref{eq:cl_src}) naively require computational cost $\bigoh(\ellmax^3)$
to compute the power spectrum for all multipoles $\ell\le\ellmax$.
For the calculations in this paper (with $\ellmax\approx 10^4$), we have found
it convenient to use position-space forms which are mathematically equivalent
but have reduced computational cost $\bigoh(\ellmax^2)$:
\ba
C_\ell^{TT ({\rm scr})} &=& 2\pi\int_{-1}^{1}d(\cos{\theta)}d^\ell_{00}(\theta)\psi^{TT}(\theta)\psi^{\tau\tau}(\theta) \, , \nn \\
C_\ell^{TE ({\rm scr})} &=& 2\pi\int_{-1}^1 d(\cos\theta)\, d^\ell_{0,-2}(\theta) \psi^{TE}(\theta) \psi^{\tau\tau}(\theta) \, , \nn \\
C_\ell^{EE ({\rm scr})} &=& \pi\int_{-1}^{1}d(\cos{\theta)}\big[d^\ell_{22}(\theta)\psi^{EE}_+(\theta)\psi^{\tau\tau}(\theta) \nn \\
&& \quad -d^\ell_{2,-2}(\theta)\psi^{EE}_-(\theta)\psi^{\tau\tau}(\theta)\big]  \, , \nn \\
C_\ell^{BB ({\rm scr})} &=& \pi\int_{-1}^{1}d(\cos{\theta)}\big[d^\ell_{22}(\theta)\psi^{EE}_+(\theta)\psi^{\tau\tau}(\theta) \nn \\
&& \quad +d^\ell_{2,-2}(\theta)\psi^{EE}_-(\theta)\psi^{\tau\tau}(\theta)\big] \,,\label{eq:cl_src_position_space}
\ea
where the correlation functions are defined by:
\ba
\psi^{\tau\tau}(\theta) &=& e^{-2\bar\tau} \sum_\ell\left({2\ell+1 \over 4\pi}\right)C_\ell^{\tau\tau}d^\ell_{00}(\theta)  \, , \nn \\
\psi^{TT}(\theta) &=& \sum_\ell\left({2\ell+1 \over 4\pi}\right)C_\ell^{TT ({\rm rec})}d^\ell_{00}(\theta)  \, , \nn \\
\psi^{TE}(\theta) &=& \sum_\ell \left( \frac{2\ell+1}{4\pi} \right) C_\ell^{TE ({\rm rec})} d_{02}^\ell(\theta) \, , \nn \\
\psi^{EE}_\pm(\theta) &=& \pm\sum_\ell\left({2\ell+1 \over 4\pi}\right)C_\ell^{EE ({\rm rec})}d^\ell_{2,\pm2}(\theta),
\ea
where $d^\ell_{2\pm2}$, $d^\ell_{0\pm2}$ and $d^\ell_{00}$ are reduced Wigner $D$-functions.
The integrals in Eqn.~(\ref{eq:cl_src_position_space}) can be done exactly, 
using Gauss-Legendre quadrature with $\lceil(3\ellmax+1)/2\rceil$ points.

\vfill
\bibliography{screening-bmode_v2}

\begin{thebibliography}{14}
\expandafter\ifx\csname natexlab\endcsname\relax\def\natexlab#1{#1}\fi
\expandafter\ifx\csname bibnamefont\endcsname\relax
  \def\bibnamefont#1{#1}\fi
\expandafter\ifx\csname bibfnamefont\endcsname\relax
  \def\bibfnamefont#1{#1}\fi
\expandafter\ifx\csname citenamefont\endcsname\relax
  \def\citenamefont#1{#1}\fi
\expandafter\ifx\csname url\endcsname\relax
  \def\url#1{\texttt{#1}}\fi
\expandafter\ifx\csname urlprefix\endcsname\relax\def\urlprefix{URL }\fi
\providecommand{\bibinfo}[2]{#2}
\providecommand{\eprint}[2][]{\url{#2}}

\bibitem[{\citenamefont{Zaldarriaga and Seljak}(1998)}]{ZalSel98}
\bibinfo{author}{\bibfnamefont{M.}~\bibnamefont{Zaldarriaga}} \bibnamefont{and}
  \bibinfo{author}{\bibfnamefont{U.}~\bibnamefont{Seljak}},
  \bibinfo{journal}{Phys. Rev.} \textbf{\bibinfo{volume}{D58}},
  \bibinfo{pages}{023003} (\bibinfo{year}{1998}), \eprint{astro-ph/9803150}.

\bibitem[{\citenamefont{Hu et~al.}(2003)\citenamefont{Hu, Hedman, and
  Zaldarriaga}}]{HuHedZal02}
\bibinfo{author}{\bibfnamefont{W.}~\bibnamefont{Hu}},
  \bibinfo{author}{\bibfnamefont{M.~M.} \bibnamefont{Hedman}},
  \bibnamefont{and}
  \bibinfo{author}{\bibfnamefont{M.}~\bibnamefont{Zaldarriaga}},
  \bibinfo{journal}{Phys. Rev.} \textbf{\bibinfo{volume}{D67}},
  \bibinfo{pages}{043004} (\bibinfo{year}{2003}), \eprint{astro-ph/0210096}.

\bibitem[{\citenamefont{Hu}(2000{\natexlab{a}})}]{Hu00}
\bibinfo{author}{\bibfnamefont{W.}~\bibnamefont{Hu}},
  \bibinfo{journal}{Astrophys. J.} \textbf{\bibinfo{volume}{529}},
  \bibinfo{pages}{12} (\bibinfo{year}{2000}{\natexlab{a}}),
  \eprint{astro-ph/9907103}.

\bibitem[{\citenamefont{{Weller}}(1999)}]{weller99}
\bibinfo{author}{\bibfnamefont{J.}~\bibnamefont{{Weller}}},
  \bibinfo{journal}{Astrophys. J.} \textbf{\bibinfo{volume}{527}},
  \bibinfo{pages}{L1} (\bibinfo{year}{1999}), \eprint{astro-ph/9908033}.

\bibitem[{\citenamefont{Liu et~al.}(2001)\citenamefont{Liu, Sugiyama, Benson,
  Lacey, and Nusser}}]{Liuetal01}
\bibinfo{author}{\bibfnamefont{G.-C.} \bibnamefont{Liu}},
  \bibinfo{author}{\bibfnamefont{N.}~\bibnamefont{Sugiyama}},
  \bibinfo{author}{\bibfnamefont{A.~J.} \bibnamefont{Benson}},
  \bibinfo{author}{\bibfnamefont{C.~G.} \bibnamefont{Lacey}}, \bibnamefont{and}
  \bibinfo{author}{\bibfnamefont{A.}~\bibnamefont{Nusser}}
  (\bibinfo{year}{2001}), \eprint{astro-ph/0101368}.

\bibitem[{\citenamefont{Santos et~al.}(2003)\citenamefont{Santos, Cooray,
  Haiman, Knox, and Ma}}]{Sanetal03}
\bibinfo{author}{\bibfnamefont{M.~G.} \bibnamefont{Santos}},
  \bibinfo{author}{\bibfnamefont{A.}~\bibnamefont{Cooray}},
  \bibinfo{author}{\bibfnamefont{Z.}~\bibnamefont{Haiman}},
  \bibinfo{author}{\bibfnamefont{L.}~\bibnamefont{Knox}}, \bibnamefont{and}
  \bibinfo{author}{\bibfnamefont{C.-P.} \bibnamefont{Ma}},
  \bibinfo{journal}{Astrophys. J.} \textbf{\bibinfo{volume}{598}},
  \bibinfo{pages}{756} (\bibinfo{year}{2003}), \eprint{astro-ph/0305471}.

\bibitem[{\citenamefont{Mortonson and Hu}(2007)}]{MorHu06}
\bibinfo{author}{\bibfnamefont{M.~J.} \bibnamefont{Mortonson}}
  \bibnamefont{and} \bibinfo{author}{\bibfnamefont{W.}~\bibnamefont{Hu}},
  \bibinfo{journal}{Astrophys. J.} \textbf{\bibinfo{volume}{657}},
  \bibinfo{pages}{1} (\bibinfo{year}{2007}), \eprint{astro-ph/0607652}.

\bibitem[{\citenamefont{Dore et~al.}(2007)}]{Doretal07}
\bibinfo{author}{\bibfnamefont{O.}~\bibnamefont{Dore}} \bibnamefont{et~al.},
  \bibinfo{journal}{Phys. Rev.} \textbf{\bibinfo{volume}{D76}},
  \bibinfo{pages}{043002} (\bibinfo{year}{2007}), \eprint{astro-ph/0701784}.

\bibitem[{\citenamefont{Hu}(2000{\natexlab{b}})}]{Hu00b}
\bibinfo{author}{\bibfnamefont{W.}~\bibnamefont{Hu}}, \bibinfo{journal}{Phys.
  Rev.} \textbf{\bibinfo{volume}{D62}}, \bibinfo{pages}{043007}
  (\bibinfo{year}{2000}{\natexlab{b}}), \eprint{astro-ph/0001303}.

\bibitem[{\citenamefont{Komatsu et~al.}(2009)}]{Komatsu:2008hk}
\bibinfo{author}{\bibfnamefont{E.}~\bibnamefont{Komatsu}} \bibnamefont{et~al.}
  (\bibinfo{collaboration}{WMAP}), \bibinfo{journal}{Astrophys. J. Suppl.}
  \textbf{\bibinfo{volume}{180}}, \bibinfo{pages}{330} (\bibinfo{year}{2009}),
  \eprint{0803.0547}.

\bibitem[{\citenamefont{Dvorkin and Smith}(2009)}]{Dvorkin:2008tf}
\bibinfo{author}{\bibfnamefont{C.}~\bibnamefont{Dvorkin}} \bibnamefont{and}
  \bibinfo{author}{\bibfnamefont{K.~M.} \bibnamefont{Smith}},
  \bibinfo{journal}{Phys. Rev.} \textbf{\bibinfo{volume}{D79}},
  \bibinfo{pages}{043003} (\bibinfo{year}{2009}), \eprint{0812.1566}.

\bibitem[{\citenamefont{Furlanetto et~al.}(2004)\citenamefont{Furlanetto,
  Zaldarriaga, and Hernquist}}]{Furlanetto:2004nh}
\bibinfo{author}{\bibfnamefont{S.}~\bibnamefont{Furlanetto}},
  \bibinfo{author}{\bibfnamefont{M.}~\bibnamefont{Zaldarriaga}},
  \bibnamefont{and}
  \bibinfo{author}{\bibfnamefont{L.}~\bibnamefont{Hernquist}},
  \bibinfo{journal}{Astrophys. J.} \textbf{\bibinfo{volume}{613}},
  \bibinfo{pages}{1} (\bibinfo{year}{2004}), \eprint{astro-ph/0403697}.

\bibitem[{\citenamefont{Zahn et~al.}(2006)}]{Zahn:2006sg}
\bibinfo{author}{\bibfnamefont{O.}~\bibnamefont{Zahn}} \bibnamefont{et~al.},
  \bibinfo{journal}{Astrophys. J.} \textbf{\bibinfo{volume}{654}},
  \bibinfo{pages}{12} (\bibinfo{year}{2006}), \eprint{astro-ph/0604177}.

\bibitem[{\citenamefont{Wang and Hu}(2006)}]{WanHu05}
\bibinfo{author}{\bibfnamefont{X.}~\bibnamefont{Wang}} \bibnamefont{and}
  \bibinfo{author}{\bibfnamefont{W.}~\bibnamefont{Hu}},
  \bibinfo{journal}{Astrophys. J.} \textbf{\bibinfo{volume}{643}},
  \bibinfo{pages}{585} (\bibinfo{year}{2006}), \eprint{astro-ph/0511141}.

\end{thebibliography}

\end{document}